\documentclass[showkeys,prl,amsmath,amssymb,floatfix,twocolumn
]{revtex4}
\usepackage{graphicx}
\usepackage{dcolumn}
\usepackage{bm}
\usepackage{natbib}
\usepackage{epstopdf}
\usepackage{natbib}
\usepackage{amsmath} 
\usepackage[abs]{overpic}
\usepackage{subfigure}
\usepackage{color} 
\usepackage{tabularx}
\newcommand{\upperRomannumeral}[1]{\uppercase\expandafter{\romannumeral#1}}

\begin{document}
\title{Curving to fly: Synthetic adaptation unveils optimal flight performance of whirling fruits}

\author{Jean Rabault, Richard A. Fauli, and Andreas Carlson}
 \email{acarlson@math.uio.no}
%
%
\affiliation{Department of Mathematics, Mechanics Division, University of Oslo.}
\date{\today}

\begin{abstract}
Appendages of seeds, fruits and other diaspores (dispersal units) are essential for their wind dispersal, as they act as wings and enable them to fly. Whirling fruits generate an auto-gyrating motion from their sepals, a leaf like structure, which curve upwards and outwards, creating a lift force that counteracts gravitational force. The link of the fruit's sepal shape to flight performance, however, is as yet unknown. We develop a theoretical model and perform experiments for double-winged bio-mimetic 3D-printed fruits, where we assume that the plant has a limited amount of energy that it can convert into a mass to build sepals and, additionally, allow them to curve. Both hydrodynamic theory and experiments involving synthetic, double-winged fruits show that to produce a maximal flight time there is an optimal fold angle for the desiccated sepals. A similar sepal fold angle is found for a wide range of whirling fruits collected in the wild, highlighting that wing curvature can aid as an efficient mechanism for wind dispersal of seeds and may improve the fitness of their producers in the context of an ecological strategy. 
\end{abstract}

\maketitle
{T}hrough evolution, plants have developed ingenious solutions for seed dispersal as a way to proliferate and to invade new territories \cite{Nathan786,bookRidley,Howe1982,Nathan2008, cain2000,green1989}. Due to their lack of mobility, they rely on mechanical principles \cite{Augspurger1986, norberg1973autorotation, lentink2009leading,Elbaum884, Marmottant20131465,Skotheim1308,Noblin1322,Armon1726} in combination with their local environment to be transported from the mother plant to a germination site. We can find plants with flight solutions for their seeds and fruits that work in ways resembling parachutes \cite{BURROWS1974, Tackenberg2003}, gliders \cite{Augspurger1986} and helicopters \cite{norberg1973autorotation,lentink2009leading}. Harnessing this aerodynamic capacity requires robust and reliable geometric designs that guide the flow of the surrounding air. 
The evolution of straight single-winged autorotating seeds indicate that flight performance has indeed been a fitness parameter \cite{Stevenson2015,Contreras2015} and these designs have spurred great interest in passive motion of solid and elastic wings \cite{smith1971,maha1999, pesavento2004, {tam2010},varshey2012, Wang2013} due to its direct relevance to the development of flight solutions. Also pollen from the genus {\em Pinus} \cite{cain1940} have evolved into shapes beneficial for their flight. 

The autogyrating motion of single bladed fruits e.g. {\em samaras} \cite{Augspurger1986, cain1940} has been studied in detail where the flow and lift production is well characterized \cite{norberg1973autorotation,lentink2009leading} in comparison to multi-winged autogyrating fruits and other diaspores, where little is know about how the shape of the wings affect their aerial motion \cite{Matlack}. The terminal descent velocity is only reported for a limited sub-set of whirling fruits, illustrated by $34$ \cite{Augspurger1986}, $53$ \cite{Tamme2014} and $16$ \cite{smith2015predicting} data entries, where the descent velocity is proportional to the square-root of the wing-loading i.e. its mass divided by the projected wing area \cite{green1989}. A complete understanding of their aerodynamics must, however, also include recordings of the rotational frequency which are elusive. To the best of our knowledge, there are no existing studies that characterize how the wing shapes of multi-winged whirling fruits influence the terminal descent velocity and their rotational frequency. When wind dispersal is the selective ecological force the growth and form of the sepals represent a subtle optimization issue for plants, because larger wings can make the seed travel greater distances. However, these are costly to build, from the perspective of energy, and can easily trap the seed in the canopy \cite{suzuki1996sepal}.


Whirling fruits also known as helicopter fruits can be found in the plant families of Dipterocarpaceae \cite{smith2015predicting,bookghazoul}, Hernandiaceae, Rubiaceae \cite{VANSTADEN1990542} and Polygonaceae, which populate Asia, Africa and the Americas, see Fig. 1a,b and include many subspecies that are currently critically endangered. For their seeds to fly long distances, the fruits are equipped with sepals, also denoted in their Greek name i.e. di = {\em two}, pteron = {\em wing} and karpos = {\em fruit}, involving an autorotating aerial motion as the fruit descends through air. Many of the tree species in these plant families are commonly known as whirling trees, helicopter trees and propeller trees. Sir David Attenborough described these winged fruits in his book on the private life of plants: ".. they are speared-shaped and curve upwards and outwards..'' \cite{nla.cat-vn610737}. Following Sir Attenborough's observation, we are intrigued by their wing geometry, which naturally leads to the following question: How does the sepal shape relate to the terminal descent velocity of the fruits?

To answer this question, we turn to a theoretical understanding of the mechanisms of flight, combined with experiments on 3D printed bio-mimetic designs of double-winged whirling fruits. We start by describing the geometrical features of these wings, which are attached to the fruit with mass $m$ and pulled to Earth by the gravitational acceleration $g$. The wings {are rigid}, have a curvature $K$, i.e. a radius of curvature $1/K$ {that is constant} along the stem, and a length $L$ that is typically much greater than the mean wing chord {along the wing} $c$, see Fig. 1a \cite{PRE}. They descend through air, which has a density $\rho$ and viscosity $\mu$, at a vertical velocity $U$ and rotate with an angular frequency $\Omega=2\pi f$ where $f$ is the rotational frequency. The wings have a pitch angle relative to the horizontal plane making the fruit auto-rotate, an action likely generated by spatiotemporal variations in the growth and drying process of the sepals \cite{BURROWS1974}. 

In order to achieve the longest flight time, the component in the vertical direction of the lift $\mathbf{F_L}$ and drag $\mathbf{F_D}$ force must be maximized and they must balance the gravitational force once the fruit reaches its terminal descent velocity $U$ (see Fig. 1c). To classify the flow, we define the Reynolds number $Re=\rho U L/\mu$, which describes the ratio of the inertial force and the viscous force. Experiments on autorotating multi-winged fruits collected in the wild fall in the regime of $Re \approx 10^3-10^4$ \cite{norberg1973autorotation,Augspurger1986, smith2016predicting} \cite{PRE}, all of which is dominated by inertia, and we know then that per wing area scales as $\vert F_L\vert \sim C_L \rho U^2$ and $\vert F_D\vert \sim C_D \rho U^2$, where $C_L$ is the lift coefficient and  $C_D$ the drag coefficient. 
To determine the optimal dispersion potential of the bio-mimetic winged fruits i.e. the minimal $U$, we place our model under an artificial evolutionary constraint by assuming that the plant has a limited energy that it can convert into a mass to build wings. We impose this constraint by fixing the wing area and keeping its thickness and chord constant. Then by performing a dimensional analysis we find, in addition to the $Re$ number, two non-dimensional numbers; the fruit's Strouhal number $St=f L/U$ giving the ratio of the rotational velocty and the translational velocity, and the fold angle $KL$. 

To understand how $U$ relates to the weight, we consider the mass flux $\rho A_d (U + U_f)/2$ through the disk area $A_d$ swept by the wings, where $U_f$ is the mean velocity of the fluid relatively to the fruit after going through the area swept by the wings. As the fluid accelerates from a rest state to a velocity of $U - U_f$ where the change in momentum must balance the weight of the fruit, i.e. $mg \approx \rho A_d (U + U_f) (U - U_f)$ \cite{norberg1973autorotation, hertel1966structure} where {$U_f \ll U$, we obtain the scaling law $U\sim\sqrt{mg / A_d}$ which also is achieved by imposing force balance between gravity and the lift force \cite{BURROWS1974}}.

The physical origin of lift and drag for $Re\gg 1$ is to be found in the details of the fluid motion \cite{lentink2009leading,sunada2015study}.
We further develop the blade element theory \cite{shapiro1956principles, norberg1973autorotation,azuma1989flight,lee2014mechanism} to describe the motion of multi-winged autorotating fruits, {which we assume descent with a constant terminal descent velocity as also confirmed in experiments}. Each wing is decomposed into a succession of thin elements across its longitudinal axis, where the relative wind velocity $\mathbf{U_r}$ is obtained as a function of the vertical descent speed $U$, the rotation frequency $\Omega$, and the local angle of attack $\alpha$. Along a wing segment the lift force ${F_L}= \frac{1}{2} \rho U_r^2 A C_L(\alpha)$ is perpendicular to the relative wind and the drag force ${F_D}= \frac{1}{2} \rho U_r^2 A C_D(\alpha)$ is tangential to the wind direction, where $A$ is the blade element area and $\alpha$ is the local angle of attack.
{We use the classical relation for $C_L(\alpha) = 2 \pi \sin(\alpha)$ \cite{landau2013fluid}, adjusted for intermediate Reynolds numbers \cite{lissaman1983low}. $C_D(\alpha)$ is parameterized from experimental measurements and our results are fairly insensitive to the choice of maximum value $\max(C_L(\alpha) / C_D(\alpha)) \in [5-13]$ before stall \cite{PRE}. Therefore, the forward ($F_F$) and normal ($F_N$)} components of the aerodynamic force on the blade element are given by

\begin{equation}
  \begin{aligned}
    F_F &= F_L \sin(\theta) - F_D \cos(\theta)\\
    F_N &= F_L \cos(\theta) + F_D \sin(\theta),
  \end{aligned}
\end{equation}

\noindent where $\theta$ is the angle between the wind direction and the horizontal plane at the location of the wing element see \cite{PRE}. The total vertical force and torque acting on the fruit are obtained by integrating along the two wingspans i.e. writing $F$ the resultant vertical force, and $M$ the moment around the rotational axis of the fruit

 \begin{equation}
 \begin{aligned}
  F &= 2 \int_{0}^{S} F_N \cos(\phi) dl -mg\\
  M &= 2 \int_{0}^{S} F_F R dl,
 \end{aligned}
 \end{equation}

\noindent where $S$ is the total curvilinear length along the wingspan, $\phi$ is the inclination of the element relative to the horizontal plane and $R$ is the distance between the wing element and the vertical axis of the fruit. {The integrals for $F$ and $M$ are computed numerically i.e. summing across all elements, where we find the terminal descent velocity when $F = M = 0$ as shown in Fig. 2a. The horizontal force on the wing can act as a motor for its rotation when the angle of the relative wind is large enough so that the forward projected component of the lift exceeds the backward component of the drag force, i.e. close to the axis of the fruit. This forward resultant force generates autorotation of multi-winged fruits, in a similar way to what is used on a helicopter \cite{shapiro1956principles}. As a consequence, a moment that drives autorotation is produced close to the axis of rotation of the fruit, while a moment that opposes rotation is produced close to the wing tips where the tangential velocity, and therefore drag, is the dominant horizontal force \cite{norberg1973autorotation,PRE}.

The rotational frequency of the fruit is determined by a balance of the total torque on the wing obtained by a projection of $\mathbf{F_L}$ and $\mathbf{F_D}$ in the horizontal plane, where the angle of attack along the wing is given by the geometry and the local relative wind direction. The ratio between rotational and translational velocity is described by the $St$ number, which naturally emerges from the expression of vertical lift ${F_L}$ and drag ${F_D}$ in the mathematical model, both scaling as $\sim U^2$ and giving $St=constant$ for a given wing geometry and is independent of fruit mass.

To challenge the theoretical predictions {that $U\sim \sqrt{mg/A_d}$} and $St=constant$ for any given fruit geometry, we perform flight measurements in a water tank using 3D printed bio-mimetic models of the double-winged fruits, {which are fully immersed in the water}. The experimental setup is designed such that the motion of these wing equipped synthetic fruits matches the Reynolds ($Re$) and Strouhal ($St$) numbers of experiments on wild fruits i.e. $Re\in[2.4-12]\times 10^3$ and $St\approx 0.4$ \cite{PRE}. By systematically increasing the weight of each fruit {while keeping its volume constant by depositing a known weight of lead to the hollow and water filled fruit}, we observe a terminal descent velocity that follows our theoretical analysis with $U\sim \sqrt{mg/A_d}$ when $A_d$ is constant. However, even taking into account that for $KL=[1.7, 3.0]$ we have a varying wing-swept area $A_d=[0.0122, 0.0116]m^2$, we obtain a clearly different descent velocity $U$ for the different fruits. 
 This illustrates that there is an addition non-trivial effect of the curved wings, which is not fully captured by the scaling law for $U$ \cite{PRE}, whereas the fruit's $St$ number is independent of the weight and found to be constant for each model. {We notice that the magnitude of $St$ in Fig. 2(b)-(d) depends only on wing geometry, where we can see that both $U$, $f$ and $St$ are non-monotonic functions of the fold angle $KL$.} Deploying our theoretical model with the constraint that wings have constant mass and only change their curvature, we ascertain an optimal wing fold angle for obtaining a maximum vertical lift force and thus best flight performance i.e. minimal $U$. {The theoretical model illustrates that although $A_d$ is nearly identical for two geometries, the vertical lift force is not uniformly distributed along the curved wings \cite{PRE}.} To further verify our hypothesis we conduct experiments on the 3D printed double-winged fruit models, where we sweep the parameter phase space in wing loadings ($mg\in[5-40]$mN) and fold angles ($KL\in[0.01-4]$) , see Fig. 2b and Fig. 3. The effect of additional parameters are further explored and discussed in \cite{PRE}, but do not qualitatively change the flight behaviour. 
 
Our man-made double-winged fruits show that to produce a maximal flight time the sepals must be curved. 
This theoretical and experimental analysis points us to a single non-dimensional parameter, the fold angle $KL$, which determines the flight performance of multi-winged helicopter fruits. {To fly with the lowest descent speed, the sepals on the fruit must be nearly horizontal at their tips to maximize the relative wind speed and the vertical lift force.} Our results point to a robust geometric criterion to maximize the lift force for this family of multi-winged fruits as shown in Fig. 3. {To see if fruits have similar shapes in the wild, we extract the fold angle $KL$ from images of a wide range of desiccated fruit wings by fitting an osculating circle along the sepal stem from its tip to its base. This gives us a measure of the radius of curvature $1/K$ along the stem and also the length $L$ of the wing i.e. the fold angle formed between the tip and the base. There are measurement errors associated with the curling of wings and that the image is not taken perpendicular to the wing. Additional measurements on synthetic fruits shows that a fruit rotated by $30^{\circ}$ gives an error of $\approx \pm 10\%$ \cite{Support}, which is illustrated by the error-bar from mean$(KL)=1.8\pm0.18$ in Fig. 3.} The measured fold angles from wild fruits are enumerated as [1 -27] in Fig. 3, and we compare directly with our theoretical predictions, which is fairly insensitive to variations in the choice of maximum $C_L/C_D$ \cite{PRE}. Through selection or chance, we see that geometries close to our optimal theoretical and experimental lift performance have evolved in nature. {Geometrical variations in the wild fruits can stem from a number of causes i.e. variation during sepal growth, different stages in their desiccation, variations among species. In ripe form many of the sepals are straight pointing along the direction of gravity and we speculate that as they lose their water content their elastic material properties become inhomogeneous, making them bend \cite{Reyssatrsif20090184,Armon1726}. Drying into shape can be a passive mechanism aiding plants to form these curvy wings to maximize flight time that can provide them with an optimal wind dispersion potential.}

A single dimensionless geometrical parameter given by the fold angle $KL$ dictates the minimal terminal descent velocity of double-winged helicopter fruits, where there is an optimal wing curvature to produce a maximal flight time. The terminal velocity of the seed is a necessary prerequisite to accurately predict their dispersion \cite{Nathan786,Tamme2014}, which can help influence policy-makers in their conservation and reforestation plans \cite{bookghazoul,smith2016predicting}. Our results may also open new ecological avenues to be explored related to their development and fitness in an ecological context. This flight strategy and mechanistic insight can have implications beyond understanding the dispersion of seeds, as they may design principles of 
flight solutions and may aid development of small scale aviation machines \cite{Ma603, ulrich2010falling}.



We are very thankful to James Smith, Jaboury Ghazoul, Yasmine Meroz, Renaud Bastien and Anneleen Kool for stimulating discussions about {\em Dipterocarpus} fruits and for their valuable input on this work. We acknowledge assistance of O. Gundersen with the experiment setup. We gratefully acknowledge financial support from the Faculty of Mathematics and Natural Science, and the UiO:LifeScience initiative at the University of Oslo.

\clearpage

\begin{figure}
  \begin{center}
    \includegraphics[width=0.5\textwidth]{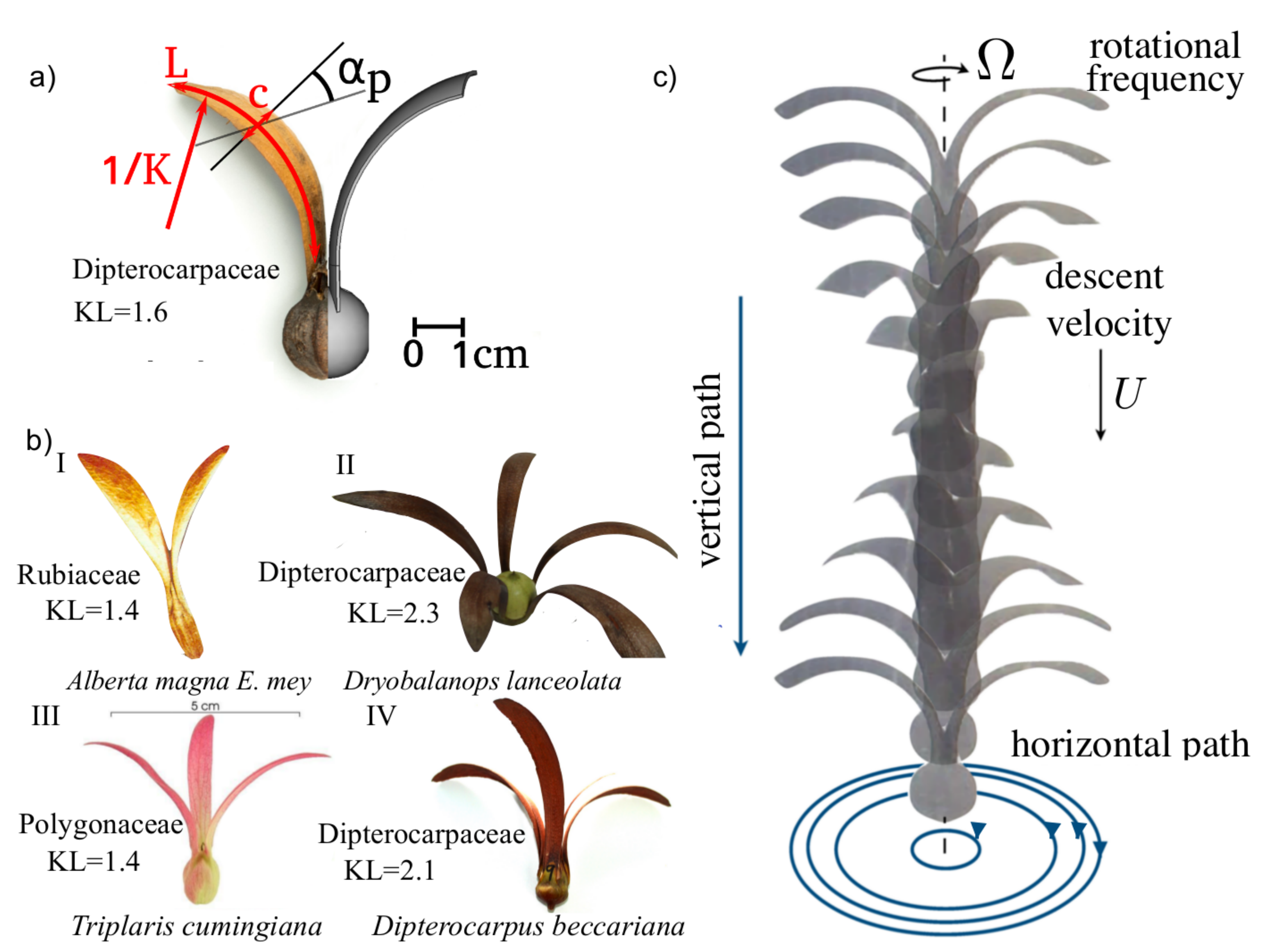}
    \caption{\label{image_dipterocarpus} a) The left part shows an illustration of the double-winged fruit of a \textcolor{red}{9}: {\it N/A} (Dipterocarpaceae) \cite{seedRef} the numbering is in correspondence with Fig. 3, where we have defined the characteristic geometrical parameters that describe its shape. We notice that the fruit is equipped with persistent and enlarged sepals of length $L$ with a curvature $K$ and thus a radius of curvature $1/K$. $c$ is the mean wing chord and $\alpha_p$ the wing angle breaking the fruits left-right symmetry. The right part of the image shows the computer aided design mimicking the helicopter fruit, which is built by the aid of 3D printing \cite{formlabs}. b) Examples of multi-winged autorotating fruits collected in the wild, illustrated by a sub-set of species; \upperRomannumeral{1}. \textcolor{red}{4}. {\it Alberta magna E. mey} ({Rubiaceae}) \cite{VANSTADEN1990542}), \upperRomannumeral{2}. \textcolor{red}{24}. {\it Shorea lanceolata} ({Dipterocarpaceae}), \upperRomannumeral{3}. \textcolor{red}{2}. {\it Triplaris cumingiana} ({Polygonaceae}) and \upperRomannumeral{4}. \textcolor{red}{22}. {\it Shorea beccariana} ({Dipterocarpaceae}) \cite{seedRef}. Images are courtesy of \cite{seedRef}. c) A schematic description of the motion of the double-winged synthetic fruit as it descends with a terminal descent velocity $U$ towards earth (vertical path) and makes one turn around the axis of rotation with a frequency $\Omega$ making it spin in a horizontal motion.
    }
 \end{center}
\end{figure}

\clearpage

\begin{figure*}
  \begin{center}
\subfigure[ ]{\includegraphics[width=.25\textwidth]{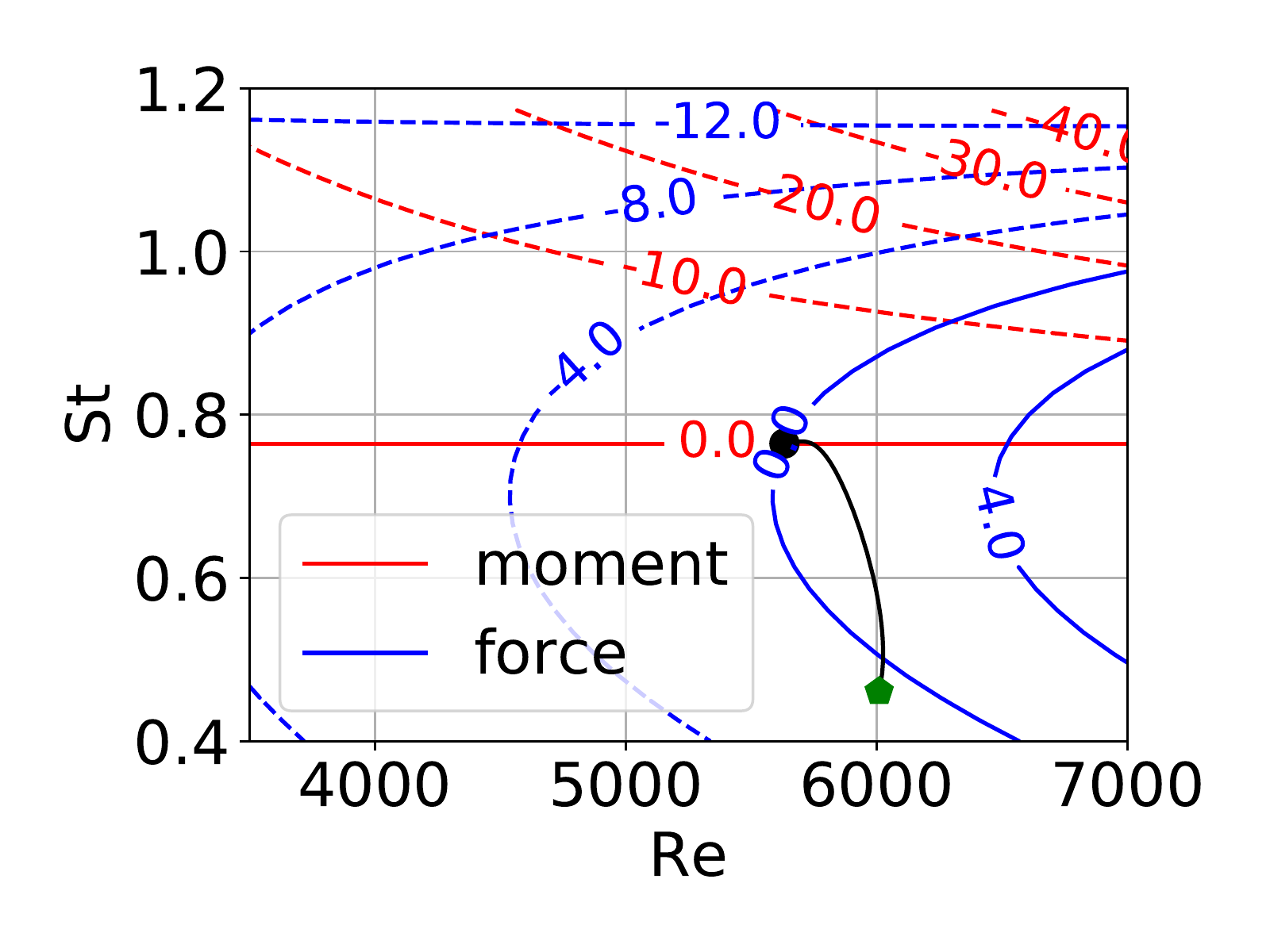}}
\subfigure[ ]{\includegraphics[width=.24\textwidth]{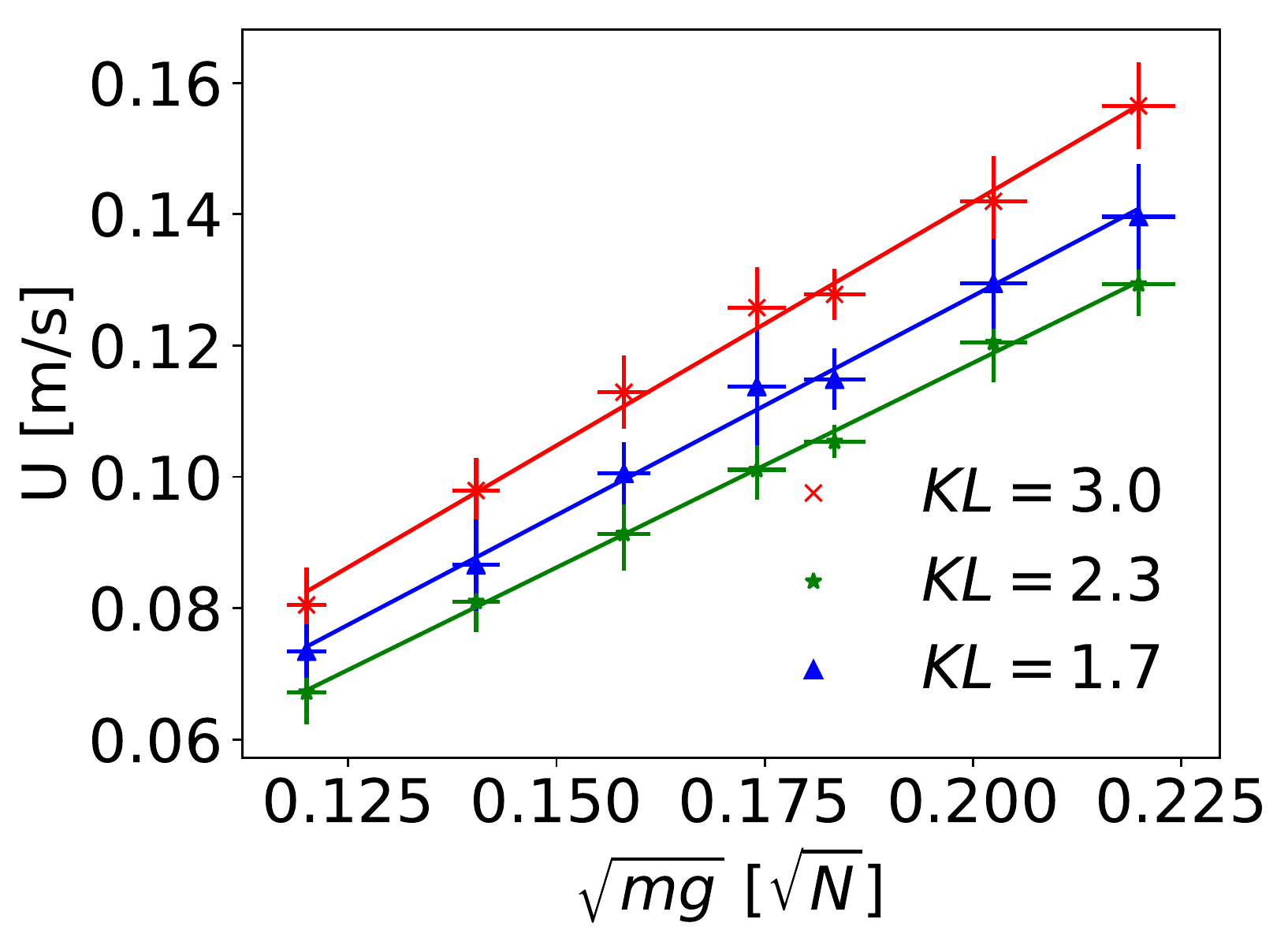}}
\subfigure[ ]{\includegraphics[width=.24\textwidth]{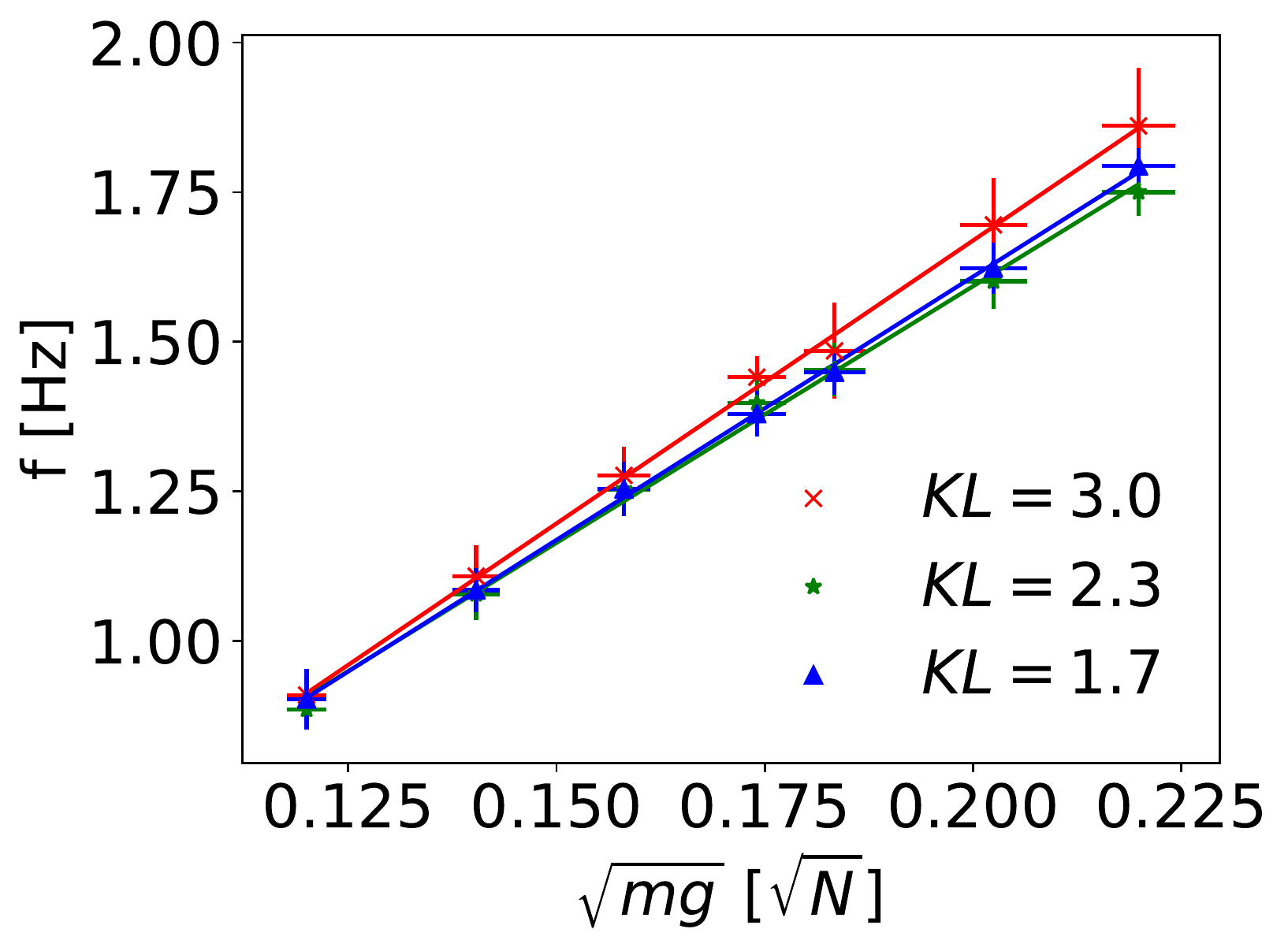}}
\subfigure[ ]{\includegraphics[width=.24\textwidth]{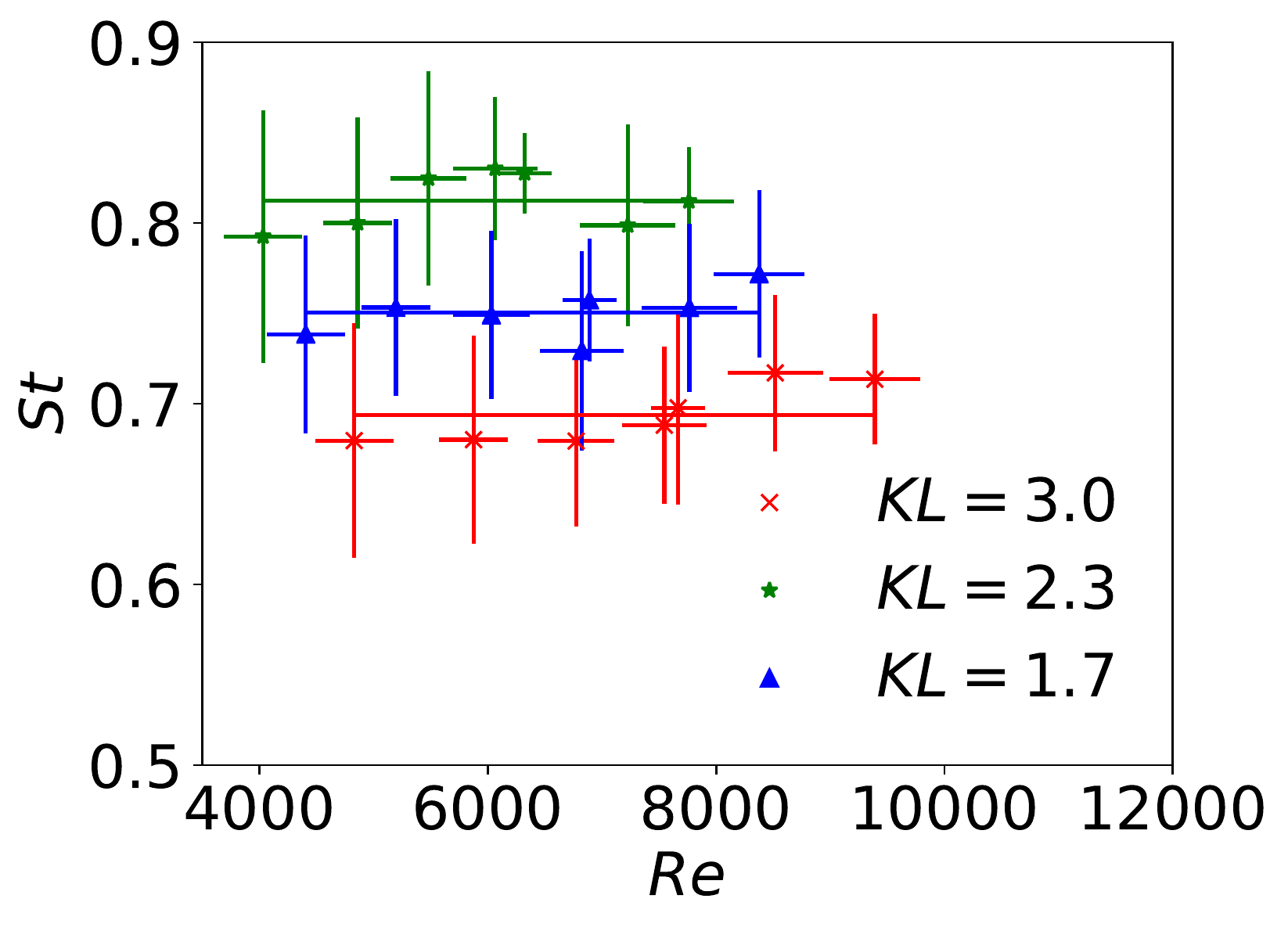}}
\caption{\label{validation_theory_wing_element} {(a) We use our mathematical blade element model to solve for the torque and vertical force balance where $KL=2.6$ and the differential mass $m = 1.2$ grams relative to the surrounding fluid. {$C_L(\alpha)$ together with a maximum value $\max{C_L(\alpha) / C_D(\alpha)} = 7$ is in direct correspondence with values expected for $Re\in 10^3-10^4$ \cite{lissaman1983low}. The fruit's predicted Reynolds ($Re$) number and the Strouhal ($St$) number are both in agreement with the experiments shown in panels (b)-(d).} The circular marker shows the point of a zero resultant force and zero moment, giving a steady vertical descent velocity $U$ and constant rotation frequency $\Omega$. A positive resultant force on the fruit would accelerate it upwards reducing its decent velocity, while a negative resultant force would increase the descent velocity of the fruit. A positive moment would generate an angular acceleration i.e. would increase the rotation rate of the fruit and vice-versa for a negative moment. In either case the fruit would return to the equilibrium position. The star-shaped marker illustrates a possible initial state and the solid line that connects to the circular marker is the path the fruit would follow. 
Measurements are made on the 3D printed bio-mimetic models where we extract; (b) terminal descent velocity $U$, (c) the rotational frequency $f$, which determine $Re$ and $St$ number in (d). 
Each data point represents the average from ten independent experiments. The error-bar in $St$ $\approx \pm 5\%$ comes from the two independent measurements of $U$ and $f$.} 
}
\end{center}

\end{figure*}

\begin{figure}
  \begin{center}
  \includegraphics[width=0.8\textwidth]{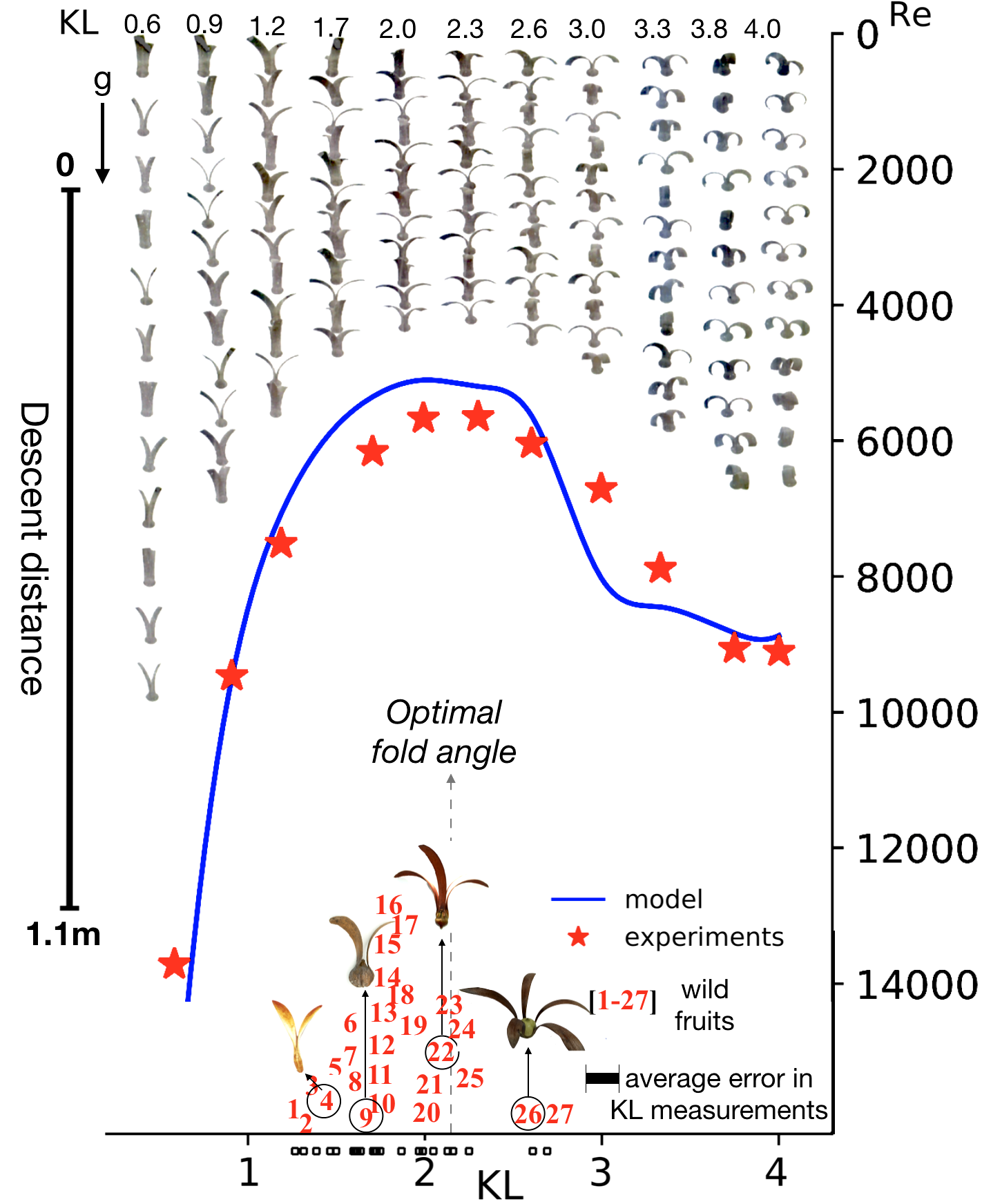}

\end{center}
\end{figure}
\begin{figure}
  \begin{center}

\caption{\label{phasespace} {We span the phase space of aerial dynamics by systematically changing the fold angle, which points to an optimal shape for production of lift force for the bio-inspired double-winged fruit. Background: a stroboscopic illustration of 11 synthetic double-winged fruits that are released at the same time and descends in the water tank, where each column corresponds to a flight duration of $5.5$ seconds. The solid line shows the vertical descent velocity as predicted by the blade element model and the star-shaped markers correspond to the experimental measurements. The theoretical model and the experimental data are in good agreement and both point us to a specific fold angle $KL$ maximizing the flight time by minimizing $U$. We show in \cite{PRE} that increasing the weight of the fruit or its wing camber only shift the curve along the y-axis (Reynolds numbers) and changing the wing angle shift the curve along the x-axis ($KL$).
The numbers above the $KL$ axis are measurements of $KL$ from desiccated sepals found in the wild \cite{seedRef}, where details about the species' and measurement procedure for $KL$ is found in the Supplemental Material \cite{Support}. 
}}
\end{center}
\end{figure}
\end{document}